\documentclass[pteplogo]{ptephy_v1} 

\preprintnumber{XXX-XXX} 

\usepackage{here}
\usepackage{bm}
\usepackage{braket}
\usepackage{graphicx}
\usepackage{caption}

\begin{document}

\title{
Pairing strength in the relativistic mean-field theory 
determined from fission barrier heights of actinide nuclei
and verified by pairing rotation and binding energies}
\author{Taiki Kouno, Chikako Ishizuka\thanks{corresponding author,\email{ishizuka.c.aa@m.titech.ac.jp}}$^*$, Tsunenori Inakura, Satoshi Chiba}
\affil{Tokyo Institute of Technology, 2-12-1 Ookayama, Meguro, Tokyo 152-8550, Japan}




%
%
%


\begin{abstract}%
\textcolor{black}{We have studied strength in the BCS pairing force,
used as a residual interaction to the relativistic mean-field \textcolor{black}{approach}, 
to reproduce height of the inner fission barriers for actinide nuclei.
}
 It was found that increasing the pairing strength \textcolor{black}{by about} \textcolor{black}{$13\%$} makes reproduce the inner fission barriers \textcolor{black}{better} over a wide range of actinide nuclei.
 This result was verified by 
using the moment-of-inertia of pairing rotational energy, \textcolor{black}{which was} 
introduced to avoid
mean-field \textcolor{black}{and odd-mass} effects in the pairing interaction to deduce purely the pairing strength.
The pairing interaction
thus determined \textcolor{black}{could simultaneously} improve \textcolor{black}{the description} of the binding energy of heavy nuclei as well.
As a result, a consistent picture among inner fission barrier, binding energy, and pairing moment of inertia \textcolor{black}{could} be obtained in terms of the relativistic mean-field
\textcolor{black}{+ BCS} theory for a broad region of the actinide nuclei.
\end{abstract}

\subjectindex{D10,D12,D29}
\maketitle

\section{Introduction}
Nuclear fission is a process where a mono-nucleus turns into two smaller fragments, occasionally \textcolor{black}{associated with emission \textcolor{black}{of} a few neutrons or light charged particles.}  This \textcolor{black}{phenomenon} is important both for application in nuclear technologies due to the large Q-values and possibility \textcolor{black}{to sustain chain reactions
mediated by neutrons}, and also for fundamental science such as  
r-process nucleosynthesis in the cosmos.  However, it is still a mysterious physics \textcolor{black}{phenomenon} as a 
large-amplitude collective motion of finite nucleon systems \cite{review}.  The first theoretical analysis was carried out by
Bohr and Wheeler, who introduced an important concept of fissility and predicted existence of fission barrier
which corresponds to the activation energy of chemical reactions \cite{Bohr}.  In their analysis, they used the liquid-drop 
model of nuclei which predicts that there is only \textcolor{black}{one} barrier, or saddle point, in the complex potential-energy 
landscape in multidimensional space of collective variables characterizing \textcolor{black}{a} nuclear shape during the fission process.
In the contemporary understanding of the nuclear fission for the actinide region where experimental data are 
most abundant, it is known that 
the nuclear fission undergoes over the fission barrier which has typically \textcolor{black}{two} humped structure, namely, over the 
inner and the outer barriers \cite{exp}.  From extensive amount of theoretical analysis, we understand now that the
inner fission barrier is located in a 
space of deformation parameters where the mass asymmetry is not important, but it is \textcolor{black}{the} case for the outer barrier. Also, it is reported that the effect of triaxiality is present for the inner barrier, though its effect on the fission barrier heights is small (typically less than 1 MeV) in the macro-micro model \cite{exp,Baran} and some microscopic \textcolor{black}{models} \textcolor{black}{\cite{RMFresult,RMFBCS,Abu,Agb,Shi}}.  However, in the calculation using \textcolor{black}{relativisitc mean field (RMF)}+point coupling model by B.N.~Lu et al. \textcolor{black}{\cite{China1,China2,China3,China4}}, it is reported that the inner barrier is reduced by about $2~{\rm{MeV}}$ by considering the \textcolor{black}{triaxiality}, and therefore this effect is noticeable.  In this manner, the effect of \textcolor{black}{triaxiality} on the inner fission barrier is highly model-dependent and its quantitative value is uncertain. Furthermore, calculating the inner fission barrier without incorporating the \textcolor{black}{triaxiality} not only saves computational time, but also makes sense since many computer codes used in applications assume axial symmetry (e.g. SkyAx \cite{SkyAx} and two-center shell model \cite{tcsm}). Therefore,we deal with the property of the 
inner fission barrier ignoring \textcolor{black}{triaxiality} as our first step of this study in this article. We understand that inclusion of \textcolor{black}{triaxiality} is definitely important for quantitative calculation of fission barriers, but it will be left as a subject of our future work.


The fission barrier has been \textcolor{black}{also} calculated by non-relativistic frameworks \cite{History,Sad}.  Besides the 
phenomenological ones whose \textcolor{black}{parameters} can be adjusted to observables, many non-relativistic microscopic 
approaches are based on Hartree-Fock theory with Skyrme \cite{Skyrme} or Gogny forces \cite{Gogny} or density-functional theory.
\textcolor{black}{Those interactions (or functionals)
are adjusted to reproduce the experimental data of ground state properties}, and have achieved great success in reproducing not only properties of the nuclear matter but
also 
binding energies, nuclear radii, \textcolor{black}{and} neutron skin thickness, over a wide range of nuclei systematically.  \textcolor{black}{The pairing interaction 
is included as a residual interaction in the BCS or Bogoliubov form.}  \textcolor{black}{In spite of their success in 
explaining ground state properties}, however, prediction of the fission barriers
remained quite poor.  In \textcolor{black}{many} cases, the fission barriers for actinides are overestimated by 2 to 5 MeV \textcolor{black}{\cite{Skyrme,Gogny}}, and 
\textcolor{black}{it has been difficult to balance the reproduction of the fission barriers and the predictions for the ground-state properties.}

The parameters describing the mean-field effects should be determined to reproduce global properties of many nuclei, so 
must not be adjusted to local property of nuclei.  On the other hand, parameters for the pairing interaction \textcolor{black}{could} be 
determined in a narrow region of nuclei, or ultimately, for nucleus by nucleus, since it describes the residual 
interaction.  Moreover, the parameters for the pairing 
interaction were determined in the past from experimental data on the gap parameters and/or even-odd staggering of nuclear 
properties \cite{RMFresult}. 
However, these methods are associated with ambiguities coming from the fact that 
1) the gap parameter and \textcolor{black}{the even-odd} staggering
are affected by the mean-field properties, and 
2) there is an ambiguity for the microscopic calculation of
odd nuclei that break the time-reversal 
symmetry.  A special treatment like blocking method is also necessary in the 
calculation for odd nuclei, but it may introduce extra sources of uncertainty. 
A novel method, therefore, should be used to determine the parameters in the pairing 
interaction not to be polluted by the mean-field effects and/or debatable method of calculation for odd nuclei.

In this work, we propose a method to determine the BCS pairing interaction strengths \cite{RMFBCS} for actinide nuclei to 
reproduce height of the inner fission barrier in the relativistic mean-field theory \cite{energy}.  Our emphasis is place to 
validate the pairing interaction thus 
determined by considering \textcolor{black}{``pairing moment-of-inertia''} \cite{pairingrotation,pairingrotation2},
which gives a property of pairing interaction avoiding \textcolor{black}{much of} the mean-field effects and \textcolor{black}{is determined} by using information only 
from \textcolor{black}{a set of} even-even nuclei.  
It has been pointed out that the pairing moment-of-inertia is an excellent experimental observable for maintaining time-reversal symmetry and measuring pair correlation properly \cite{pairingrotation,pairingrotation2}.
\textcolor{black}{In addition, we verify that the pairing interaction determined here can provide a better prediction for the ground state binding energies, \textcolor{black}{pairing moment-of-inertia} 
and fission barrier heights simultaneously.}

Our paper is organized in the following manner. In sect.~2 we introduces the relativistic mean-field theory, pair correlation, and pairing rotation energy.  
In subsect.~3.1, we will first look at the changes in the inner fission barrier when the pair correlation force is changed, taking $^{240}{\rm Pu}$ as an example, and evaluate the change in the pair rotation energy according to the change of the pair correlation force.  
In subsect.~3.2, the scope of investitaion is expanded to a broader region of actinide nuclei, and the pair correlation force that reproduces the
experimental inner-fission barriers is obtained \textcolor{black}{by referring to the Reference Input Parameter Library (RIPL-3) \cite{RIPL3}.  This is a highly important database for theoreticians involved in the development and use of nuclear reaction modeling (ALICE \cite{ALICE}, EMPIRE \cite{EMPIRE}, GNASH \cite{GNASH},
UNF \cite{UNF}, TALYS \cite{TALYS}, CCONE \cite{ccone}, and so on) both for theoretical research and nuclear data evaluations.  In addition, RIPL3 is often used to compare experimental and 
theoretical values of wide variety of nuclear quantities.}
Furthermore, by using this pair correlation force, we verify that the binding energy and also the pair rotation are
described simultaneously better than the original pairing interaction.  In other words, \textcolor{black}{we have shown} that there is a possibility that the pair rotation 
\textcolor{black}{is} used to determine the pairing strength in a systematical manner.

\section{Method}
\subsection{Relativistic Mean Field theory}
We start with the following Lagrangian density considering relativistic invariance
%
\begin{eqnarray}
\mathcal{L}_{\rm RMF}
&=&\bar{\psi}(i\gamma_{\mu}\partial^{\mu}-M)\psi+\frac{1}{2}\partial^{\mu}\sigma\partial_{\mu}\sigma-U(\sigma)-g_{\sigma}\bar{\psi}\psi\sigma \nonumber\\
& &-\frac{1}{4}\Omega^{\mu\nu}\Omega_{\mu\nu}+\frac{1}{2}m_{\omega}^{2}\omega^{\mu}\omega_{\mu} 
-g_{\omega}\bar{\psi}\gamma^{\mu}\psi\omega_{\mu}+U(\omega) \nonumber\\
& & -\frac{1}{4}{\bm{R}}^{\mu\nu}\cdot{\bm{R}}_{\mu\nu} 
+\frac{1}{2}m_{\rho}^{2}{\bm{\rho}}^{\mu}\cdot{\bm{\rho}}_{\mu}-g_{\rho}\bar{\psi}\gamma^{\mu}{\bm{\tau}}\psi{\bm{\rho}}_{\mu} \nonumber\\ 
& &-\frac{1}{4}F^{\mu\nu}F_{\mu\nu}-e\bar{\psi}\gamma^{\mu}\frac{(1-\tau_{3})}{2}\psi A_{\mu}, \\
\Omega^{\mu\nu}&=&\partial^{\mu}\omega^{\nu}-\partial^{\nu}\omega^{\mu},\ {\bm{R}}^{\mu\nu}=\partial^{\mu}{\bm{\rho}}^{\nu}-\partial^{\nu}{\bm{\rho}}^{\mu},\  F^{\mu\nu}=\partial^{\mu}A^{\nu}-\partial^{\nu}A^{\mu}
\end{eqnarray}
including the non-linear terms 
\begin{eqnarray}
U(\sigma)&=&\frac{1}{2}m_{\sigma}^{2}\sigma^{2}+\frac{1}{3}b_{2}\sigma^{3}+\frac{1}{4}b_{3}\sigma^{4},\ \\
U(\omega)&=&\frac{1}{4}c_{3}(\omega_{\mu}\omega^{\mu})^{2} 
\end{eqnarray}
as self-interaction
terms \cite{Lagrangian1,Lagrangian2,Para}. 
Here, \textcolor{black}{$M$ is the nucleon mass, $m_{\sigma},m_{\omega},m_{\rho}$ denote the meson masses, and $g_{\sigma},g_{\omega},g_{\rho}$ represent the meson-nucleon coupling constants. Furthermore, $\sigma$, $\omega^{\mu}$, ${\bm{\rho}}^{\mu}$ 
and $A^{\mu}$ indicate 
\textcolor{black}{scalar-isoscalar field},  \textcolor{black}{vector-isoscalar field},  \textcolor{black}{vector-isovector field}
and the photon field, respectively. The symbol $\psi$ represents the nucleon Dirac field consisting of four components.} 
In the relativistic mean-field approach, however, \textcolor{black}{the}
fluctuation of the meson field is ignored as well as ignoring the negative energy components (so-called no-sea approximation). 
Furthermore, besides the non-linear self-interaction term of the \textcolor{black}{$\sigma-$}meson which has been commonly used, 
the non-linear self-interaction term of the \textcolor{black}{$\omega-$}meson, whose importance has been suggested by Relativistic Brueckner-Hartree-Fock (RBHF) theory in recent years \cite{Lagrangian2}, is included.  For this sake, we used \textcolor{black}{the} parameter set denoted as NLV-20 \cite{Para} \textcolor{black}{primarily}, 
\textcolor{black}{although the conclusion obtained in this work does not depend on
particular choice of parameter set as long as we investigated.}
\begin{table}[H]
  \begin{center}
    \caption{\textcolor{black}{Parameter values used in NLV-20. \textcolor{black}{$M,m_{\sigma},m_{\omega}$,and $m_{\rho}$ are in units of ${\rm{MeV}}$
    and $b_{2}$ is in units of ${\rm{fm}}^{-1}$. The other parameters are dimensionless.}}}
    \begin{tabular}{|l|l|l|l|l|l|l|l|l|l|l|} \hline   
$M$ & $m_{\sigma}$ &$m_{\omega}$&$m_{\rho}$&$g_{\sigma}$&$g_{\omega}$&$g_{\rho}$&$b_{2}$&$b_{3}$&$c_{3}$  \\ \hline
$938.9$& $489.049$& $780.0$&$763$&$10.0518$&$12.9354$&$4.90748$&$-12.7384$&$-34.0567$&$20.0$ \\ \hline
    \end{tabular}
  \end{center}
\end{table}
\textcolor{black}{The parameters of NLV-20 are shown in Table 1 that has been adjusted to reproduce the binding energies of $^{16}{\rm{O}}$, $^{40,48}{\rm{Ca}}$, $^{56,58}{\rm{Ni}}$, $^{88}{\rm{Sr}}$, $^{90}{\rm{Zr}}$,
$^{112,124,132}{\rm{Sn}}$, $^{146}{\rm{Gd}}$ and $^{208}{\rm{Pb}}$, the diffraction radii of $^{16}{\rm{O}}$, $^{40,48}{\rm{Ca}}$, $^{56,58}{\rm{Ni}}$, $^{88}{\rm{Sr}}$, $^{90}{\rm{Zr}}$, $^{112,124}{\rm{Sn}}$, $^{146}{\rm{Gd}}$ and $^{208}{\rm{Pb}}$, and surface thicknesses of $^{16}{\rm{O}},^{40,48}{\rm{Ca}},^{90}{\rm{Zr}},^{112,124}{\rm{Sn}}$ and $^{208}{\rm{Pb}}$.  \textcolor{black}{The saturation density obtained with this parameter set is $0.151~\rm fm^{-3}$}, the binding energy per nucleon is $16.24~\rm MeV$, the \textcolor{black}{incompressibility} is $190~\rm MeV$, and the symmetry energy per nucleon is $42.1~\rm MeV$ \textcolor{black}{for the symmetric nuclear matter}.  Please note that the parameters were 
determined by properties of spherical nuclei \textcolor{black}{alone} ranging from $^{16}$O to $^{208}$Pb.}

\textcolor{black}{We used the Nilsson oscillators in the basis expansion and cylindrical coordinate for the calculations with NLV-20, and deformed harmonic oscillators in the basis in the DD-ME2 and DD-PC1 cases. As for the numerical details, please have a look at \cite{Para,program}}.

The total energy was calculated by volume-integrating the energy density obtained as the $(00)$ component of the energy-momentum tensor \cite{energy}. Then, corrections were given for the center-of-mass  energy \textcolor{black}{$E_{\rm cm}=\Braket{\hat{P}^{2}_{\rm cm}}/2AM$, where \textcolor{black}{$\hat{P}_{cm}$ is the center-of mass momentum operator \cite{Para}}},
and the pairing energy to be explained in the next subsection.
In calculating the total energy as a function of deformation to cover the inner fission barrier, we imposed a constraint on the quadrupole moment calculated from the nucleon density \textcolor{black}{$\rho({\bm r})$}:
\begin{eqnarray}
Q_{20}=\frac{1}{2}\sqrt{\frac{5}{4\pi}}\int d^{3} {\bm{r}} \rho({\bm{r}})(2z^{2}-x^{2}-y^{2})
\end{eqnarray}

\subsection{Pairing interaction}

The effective pairing interaction is parametrized by using a local pairing energy functional as follows \cite{bender}:
\begin{equation}
\varepsilon_{pair}=\frac{1}{4} \sum_{q \in \{p,n\}} \int d^3 {\bm r} \chi_{q}^*({\bm r}) \chi_q({\bm r}) G_q({\bm r}),   
\end{equation}
where $\chi_q({\bf r})$ is the local part of the pair density matrix obtained as:
\textcolor{black}{\begin{eqnarray}
\chi_q({\bm r})& = &-2 \sum_{k>0}f_{k} u_{k,q} v_{k,q} \left | \phi_{k,q}({\bf r}) \right |^2,\\
f_{k}&=&\frac{1}{1+\exp[(\epsilon_{k}-\lambda_{q}-\Delta E_{q})/\mu_{q}]}
\end{eqnarray}}
with $\phi_{k,q}$ the single particle wave function for $q \in \{p,n\}$. The symbols
$ u_{k,q}$ and  $v_{k,q}$ denote, respectively, the vacant and occupied amplitudes
of a single orbit $k$ obtained by the BCS theory. \textcolor{black}{Furthermore, $f_{k}$ is the energy-dependent cutoff weights, $\Delta E_{q}$ and $\mu_{q}=\Delta E_{q}/10$ denotes cutoff parameters \cite{bender}.
}
In the present work, we used a simple constant pairing strength as $G_q({\bf r}) = 
G_q$
corresponding to the \textcolor{black}{delta pairing interaction. The pairing strengths are taken from \cite{Para}, $G_{n}=-348~{\rm{MeV\ fm^{3}}}$ for neutrons and $G_{p}=-349~{\rm{MeV\ fm^{3}}}$ for protons.} 
These values \textcolor{black}{are} determined as follows. \textcolor{black}{In \cite{Para}, the pairing strengths were determined by reproducing the experimental pairing gaps of $^{44}{\rm{Ca}}$, $^{106}{\rm{Sn}}-^{128}{\rm{Sn}}$, and $^{201}{\rm{Pb}}-^{204}{\rm{Pb}}$ for neutrons and of $^{52}{\rm{Cr}}$, $^{82}_{50}{\rm{Ge}}-^{94}_{50}{\rm{Ru}}$, $^{136}_{82}{\rm{Xe}}-^{147}_{82}{\rm{Tb}}$, and $^{212}_{126}{\rm{Rn}}-^{215}_{126}{\rm{Ac}}$ for protons.}
\textcolor{black}{Hereafter, a set of these numbers is denoted as $G=(-348,-349)$ \textcolor{black}{\cite{Para}}.}
%

\subsection{Pairing rotation}
In this work, we adjust the strength of the pairing interaction to reproduce the height of the inner fission barriers of nuclei in the actinide region.  Such a procedure can be justified only if the \textcolor{black}{pairing} strength thus determined is firmly consistent with the quantity related to the pairing correlation.  
Normally, the evaluation of \textcolor{black}{the} pair correlation is performed using the \textcolor{black}{pairing} gap and/or odd-even staggering of binding energy or neutron separation energy. \textcolor{black}{But} this method suffers from \textcolor{black}{the} effects coming \textcolor{black}{from} the mean-field part, and also is difficult since it involves calculation for odd nuclei that break the time-reversal symmetry. Numerically, a special technique like "blocking method" must be employed for calculation of odd nuclei, which introduce extra ambiguity.  Instead,  we bring a concept called moment-of-inertial of ``pairing rotation" \textcolor{black}{\cite{pairingrotation,pairingrotation2}}.  \textcolor{black}{This quantity can  be evaluated by using the binding energy of even-even nuclei alone}, and hence is an excellent experimental observable for obtaining information on pair correlations that maintain time-reversal symmetry, while reducing the mean-field effects simultaneously.

The pairing rotation energy is defined by the following equation, derived by the pair correlation breaking the \textcolor{black}{U(1)} gauge symmetry \cite{pairingrotation,pairingrotation2}:
\textcolor{black}{\begin{eqnarray}
E(N,Z_{0})=E(N_{0},Z_{0})+\lambda_{n}(N_{0},Z_{0})\Delta N+\frac{(\Delta N)^{2}}{2\mathcal{I}_{nn}(N_{0},Z_{0})
\label{pr1}}
\end{eqnarray}}
where \textcolor{black}{$E(N_0,Z_0)$ denotes ground-state energy} for neutron number of $N_0$ and proton number of $Z_0$ nucleus, $\Delta N=N-N_{0},\ \lambda(N_{0})=dE/dN|_{N=N_{0}}$ is the chemical potential, \textcolor{black}{and the second-order term is the pairing-rotational energy with the \textcolor{black}{pairing moment-of-inertia} $\mathcal{I}_{nn}(N_0,Z_0)^{-1}=d^{2}E/dN^{2}|_{N=N_{0}}$.} The pairing moment-of-inertia is given by the reciprocal of the second order derivative of the energy around ($N_0,Z_0$) nucleus which can be calculated by the following \textcolor{black}{finite differentiation}:
\begin{eqnarray}
\mathcal{I}_{nn}^{-1}(N_0,Z_0)=\frac{E(N_0+2i,Z_0)+E(N_0-2i,Z_0)-2E(N_0,Z_0)}{4i}, 
\label{pr2}
\end{eqnarray}
where $i=1$ is selected \textcolor{black}{usually} to  calculate the second derivative of the pairing rotation energy by using nuclei having \textcolor{black}{$N_0-2$, $N_0$ and $N_0+2$}, but for some cases $i=2$ gives a 
better description of the second-order derivative, eq. (\ref{pr1}), or to discriminate different 
parametrizations.
 This quantity can be calculated by using the binding energy of even-even nuclei where the experimental values were taken from ``AME2016" \cite{bindexp}.
\section{Results and Discussion}

\subsection{Total Energy}

\textcolor{black}{In this paper, we examine the influence of the pairing rotation for six actinides ($^{234}{\rm{U}}$,$^{236}{\rm{U}}$,$^{240}{\rm{Pu}}$,$^{242}{\rm{Pu}}$,$^{242}{\rm{Cm}}$,$^{244}{\rm{Cm}}$) which are \textcolor{black}{representative} compound nuclei synthesized in neutron-induced reactions.}

\textcolor{black}{Firstly, we show the dependence of the total energy \textcolor{black}{of} $^{240}{\rm{Pu}}$ on quadrupole moment in three cases where the pairing strength $0.8G,G,1.2G$. Here $G$ is the original value $G=(G_{n},G_{p})=(-348,-349)$ MeV \textcolor{black}{fm$^3$}.} 
\textcolor{black}{Since the original value $G$ was determined from even-odd mass  staggering which has ambiguity in connection with the \textcolor{black}{pairing} strength, we adjust the \textcolor{black}{pairing} strength to reproduce the heights of inner fission barrier, and validated the results by other quantities, namely moment-of-inertia of pairing rotation and binding energies. Furthermore, there is a possibility to adjust the $G_n$ and $G_p$
independently.  However, we did not dare to do that since we tried 
to minimize the extra freedom we introduced in this work.}

\begin{figure}[htbp]
\centering\includegraphics[scale=0.6]{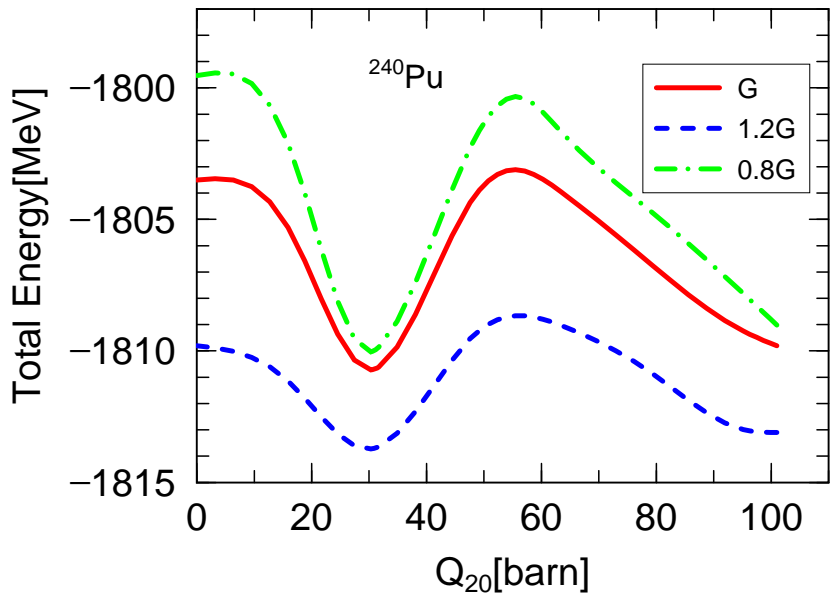}
\caption{Dependence of total energy of $^{240} {\rm{Pu}}$ on the \textcolor{black}{quadrupole moment} for \textcolor{black}{three} values of pairing strength parameters.}
\label{Fig1}
\end{figure}

\textcolor{black}{Fig.~\ref{Fig1} shows how the total energy of $^{240}$Pu depends on $Q_{20}$ for 3 sets of the pairing strengths.}  It is seen from Fig.~\ref{Fig1} that the position of the ground-state stays at $Q_{20}\simeq30~{\rm{barn}}$, while the top of the inner barrier, namely, saddle point, stays at $Q_{20}\simeq60~{\rm{barn}}$ regardless of the strength of the pairing interaction. The height of the inner fission barrier was obtained as \textcolor{black}{the difference between the energies of the saddle point and the ground state.}
\begin{figure}[htbp]
\centering\includegraphics[scale=0.8]{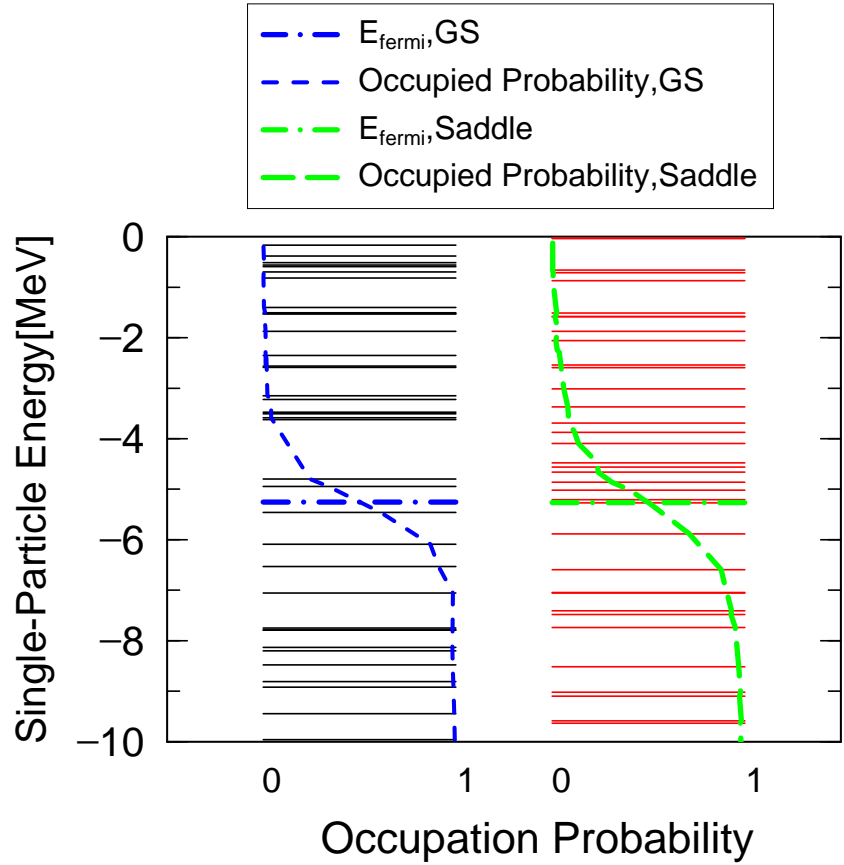}
\caption{Neutron single-particle energies of $^{240}{\rm{Pu}}$ at \textcolor{black}{the ground state} (left) and saddle point (right) . In addition, the broken lines represent occupied probabilities at both points.  The horizontal \textcolor{black}{dot-dashed lines} denote the Fermi energy (or chemical potential in BCS theory). \textcolor{black}{Here "GS" is "Ground States".}}

\label{Fig3}
\end{figure}
\begin{figure}[htbp]
\centering\includegraphics[scale=0.45]{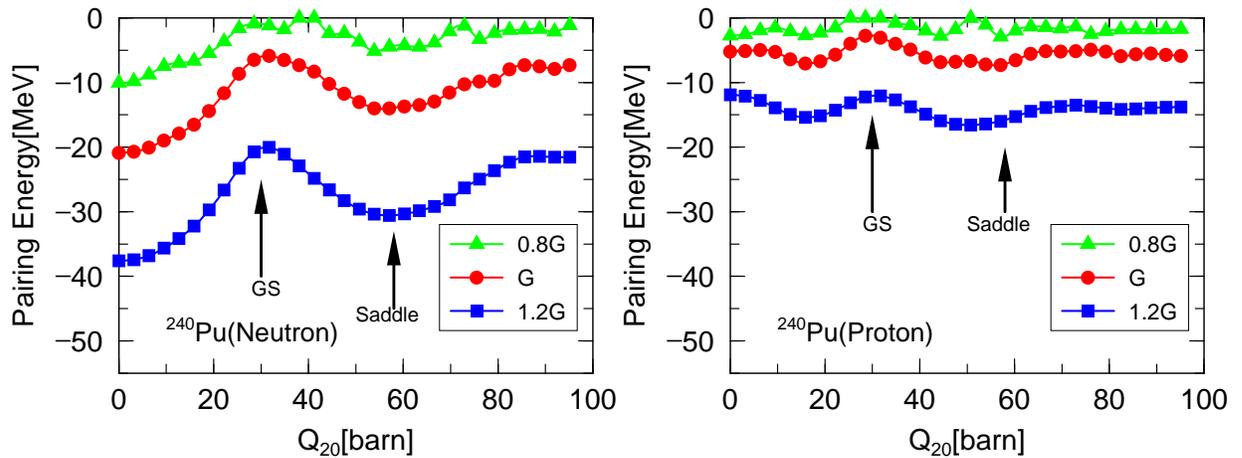}
\caption{Pairing Energy of $^{240}{\rm{Pu}}$ as function of the deformation by each pairing strength. Left panel: the pairing energy of neutron, right panel: the pairing energy of proton.}
\label{Fig12}
\end{figure}
\begin{figure}[hbp]
\centering\includegraphics[scale=0.7]{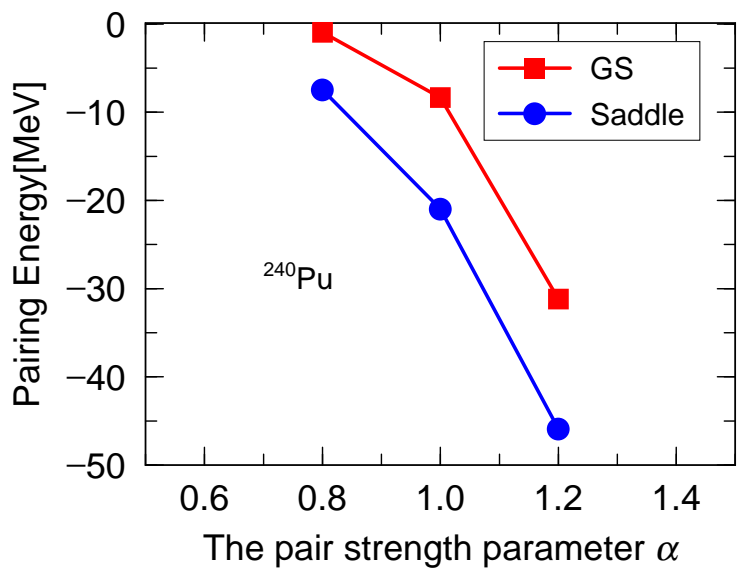}
\caption{The changes of pairing energy by each pair strength on ground states and saddle point.}
\label{Fig14}
\end{figure}
The experimental value of the height of the inner fission barrier, taken from RIPL3\cite{RIPL3}, is $6.05 ~\rm{MeV}$, while these three values of the pairing strength \textcolor{black}{yield} the following three results:
\begin{eqnarray}
G\rightarrow7.54~{\rm{MeV}},\ 1.2G\rightarrow5.08~{\rm{MeV}},\ 0.8G\rightarrow9.71~{\rm{MeV}}.
\end{eqnarray}
 \textcolor{black}
 {Hereafter, we adjust the pairing strengths of both neutrons and protons by a single multiplier parameter $\alpha$ as $G \to \alpha G$.  One of the reasons} why the height of the inner fission barrier changes depending on the strength of the pairing force is due to the fact that the single-particle level densities around the Fermi surface are different at
 the ground state and the saddle point. \textcolor{black}{This physical picture is understood by Fig. 2, which depicts neutron single-particle energies of $^{240}{\rm{Pu}}$ at the ground state and saddle point.} 
 \textcolor{black}{Furthermore, from Fig. 3, which shows the deformation dependence of the pairing energy at each pairing strength, it is possible to know how the pairing energy works at each degree of deformation.}
 In particular, these values are maximal at the \textcolor{black}{GS} (Ground State) while they take minimal values at the saddle point. Such physical picture can be understood by the following equation proposed by Nilsson and Ragnarsson \cite{Nillson} intuitively, which defines the \textcolor{black}{difference between the energies of} correlated state $ E(\Delta \neq0) $ and the unpaired state $ E(\Delta=0)$:
\begin{eqnarray}
E(\Delta)-E(0)\simeq-\frac{1}{2}\rho\Delta^{2}
\label{eq:nils1}
\end{eqnarray}
where $\Delta$ is BCS-gap parameter and $\rho$ is level density.
\textcolor{black}{The gap $\Delta$ is also given approximately as 
\begin{equation}
    \Delta \sim 2S \exp \left ( -\frac{1}{G_0 \rho} \right ),
\label{eq:nils2}
\end{equation}
which indicates that it increases as $\rho$ increases.  Here, 
$S$ denotes an energy range where the $u$ and $v$ factors deviate
noticeably from 0 and 1, respectively, }\textcolor{black}{and $G_0$ denotes the pairing strength in MeV when the 
pairing interaction is written in the form of 
$H_{pair}=G_0 \sum_\mu \sum_\nu a_\mu^\dag  a_{\bar{\mu}}^\dag a_{\bar{\nu}} a_\nu $ \cite{Nillson}.}
From \textcolor{black}{Eqs.~(\ref{eq:nils1}) and (\ref{eq:nils2})} and Figs. \ref{Fig3} and Fig. \ref{Fig12}, it can be seen that the pairing effect is stronger as a negative value at the saddle point than in the ground state since the level density is higher in the former. \textcolor{black}{We also notice that the change of the pairing energy is stronger for neutrons than for protons.} Furthermore, as shown in Fig.~\ref{Fig14}, \textcolor{black}{which depicts the changes of pairing energy by each pairing strength on ground states and saddle point,} it can be seen that the slope of the pairing energy dependent on the \textcolor{black}{pairing} strength parameter $\alpha$ is steeper at the saddle point where the level density is higher. 
%
%
It is concluded that the fission barrier changes by these effects  of pairing correlation \textcolor{black}{as one of the typical reasons}. The above discussion was conducted with reference to the paper written by S.~Karatzikos et al. In addition, a more detailed analysis and discussion is also given in \cite{RMFBCS}. 
 
\begin{figure}[htbp]
\centering\includegraphics[scale=0.6]{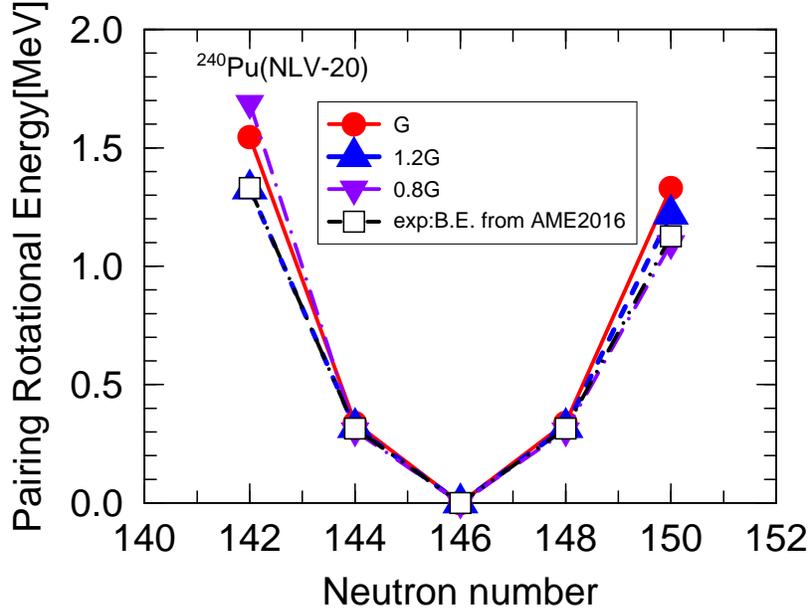}
\caption{Pairing rotation energy by changing the pairing strength in \textcolor{black}{$^{240} {\rm{Pu}}$} with 
NLV-20.}
 \label{Fig4}
\end{figure}

 %
 In this example, increasing the value of pairing strength decreases the overall
 energy of the system, while change is smaller at the ground state than at the 
 saddle point.  \textcolor{black}{Accordingly, it} changes the height of the 
 inner fission barrier, and agreement to the experimental value becomes better.  The
 crucial point here is, however, that if the increase of the pairing strength gives 
 a consistent picture, or better reproduction, for the pairing rotation simultaneously or not.  
To see this, we plot the pairing rotation
 energy in Fig.~\ref{Fig4}.  
It can be seen that the experimental value of pairing rotational energy around 
$^{240}$Pu is reproduced almost equivalently when the pairing strength is increased
 with the original pairing 
strength for $i=1$ (see eq. (\ref{pr2})), or better  
 for $i=2$ case. In particular, it can be seen that the result of $0.8G$ and $G$ at $N=142$ is significantly worse than $1.2G$.
 The deviation between the calculated pairing-rotational energy at $N=150$ seems to be smallest with $0.8G$. However, the total deviation in $N=142-150$ becomes smallest with $1.2G$.
Similar plots were made for 
 different interactions, \textcolor{black}{DD-ME2} \cite{DDME2} and \textcolor{black}{DD-PC1} \cite{DDPC1} in Fig.~\ref{DDME2andDDPC1}.  In the \textcolor{black}{latter two interactions}, agreement with the 
 measured pairing rotational energy is drastically improved when the pairing strength is increased by \textcolor{black}{20\%} compared with the original strength and the case where the pairing strength is reduced by \textcolor{black}{20\%}.  The values of fission barriers by this two latter,  density-dependent, interactions with the triaxiality taken into account are 7.50 and 6.53 MeV, respectively, which are still 1 to 2 MeV too large compared to the measured value.  \textcolor{black}{This is the same situation as calculation with NLV-20 interaction, without triaxiality.}
This \textcolor{black}{result indicates} the procedure to 
increase the pairing strength in the actinide region to reproduce values of the inner fission barrier 
better.  Therefore, we will proceed to perform a systematical analysis in this direction
by adopting NLV-20 interaction in the followings.

\subsection{Pairing strength}
%
%
\begin{figure}[htbp]
\centering\includegraphics[scale=0.45]{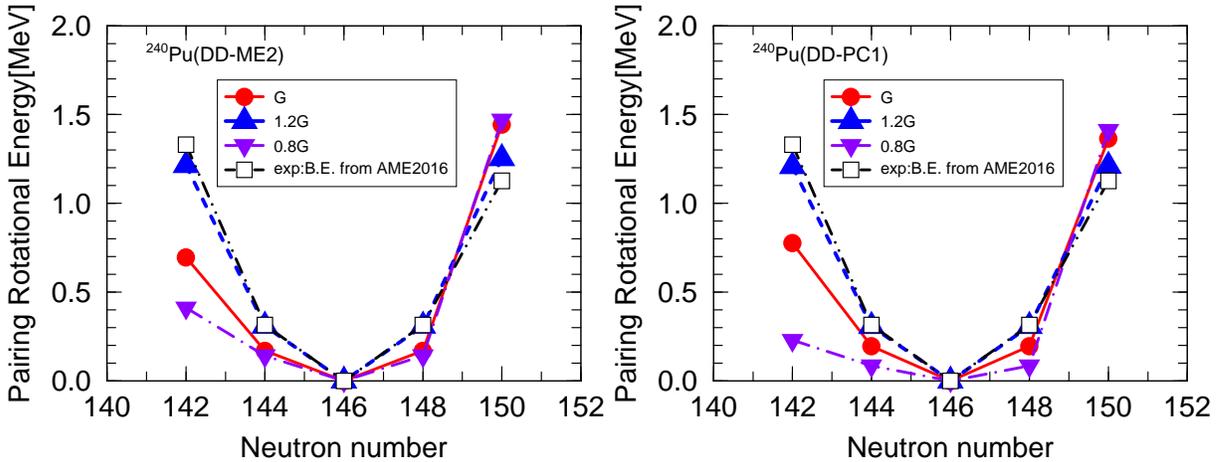}
\vspace{-0.5cm}
\caption{Pairing rotation energy by changing the pairing strength in \textcolor{black}{$^{240} {\rm{Pu}}$}
with \textcolor{black}{DD-ME2} \cite{program,DDME2} (left figure) and \textcolor{black}{DD-PC1} \cite{program,DDPC1} (right figure). \textcolor{black}{Lines are drawn only as eye-guides, and only values at the 
even integer values of the neutron number have meanings.}}
 \label{DDME2andDDPC1}
\end{figure}

From the preceeding analysis, we found that the fission barrier can be adjusted by changing the pairing strength, while making agreement to the pairing rotational energy much better than using the original pairing strength as shown in Figs.~\ref{Fig4} and    \ref{DDME2andDDPC1}.
Therefore, we selected \textcolor{black}{six} nuclei ($^{234}{\rm{U}},^{236}{\rm{U}},^{240}{\rm{Pu}},^{242}{\rm{Pu}},^{242}{\rm{Cm}},^{244}{\rm{Cm}}$) from the actinide region for which the moment-of-inertia of the pairing rotation can be defined well, and determined the pair correlation force that reproduces the experimental values of the inner fission barriers taken from RIPL3\cite{RIPL3}.  Such a pairing strengths adjusted to reproduce the inner fission barrier is denoted as $G_{\rm best} \equiv \alpha \cdot G$ and the pair strength parameter $\alpha$ are 
\textcolor{black}{shown in Fig. 7 as a function of $Z^{2}/A^{1/3}$ of the fissioning nuclei.  A straight line assuming a weak dependence on $Z^{2}/A^{1/3}$ is drawn, but it is 
almost equivalent with the average $\alpha$ value of $1.127$.}
%
%
%
\begin{figure}[htbp]
\centering\includegraphics[width=3.2in]{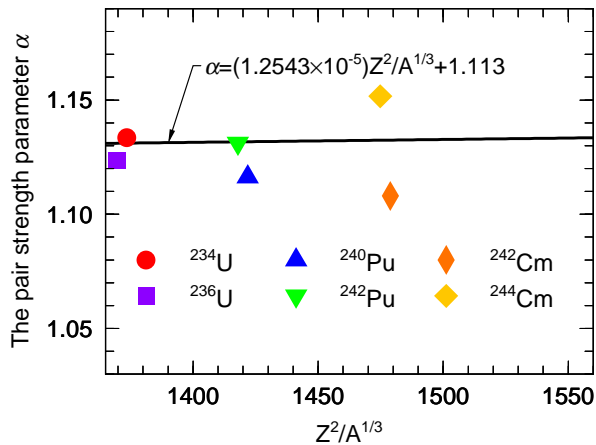}
\caption{The pair strength parameter $\alpha$ with $Z^{2}/A^{1/3}$. 
The solid line is a linear fit $\alpha$, and $G_{{\rm best}} = \alpha (\mathrm{for~each~nucleus)}) \cdot G$.}
\label{Fig5}
\end{figure}
%

\textcolor{black}{We calculated the root-mean square (RMS) error by
\begin{eqnarray}
RMS=\sqrt{\sum_{i=1}^{6}\frac{(x_{i}(th)-x_{i}(exp))^{2}}{6}},~ 
\{x_{i}\equiv \mbox{$B_f$},\mbox{Binding Energy},\mathcal{I}_{nn} \mbox{ for $i$-th nucleus}\} 
\end{eqnarray}} 
\textcolor{black}{where $x_i(th)$ and $x_i(exp)$ denotes theoretical and experimental values,  respectively.}  
Table 2 summarizes the RMS errors of the \textcolor{black}{three} quantities depending on the pairing force strength.  It should be noticed that the binding energy calculated with the $ G_ {\rm best} $ becomes better for all the considered nuclides: the RMS error is improved by about $50\%$. Similarly, it can be seen that the pairing moment-of-inertia is also improved by calculation with $ G_{\rm best}$,
resulting that the RMS error is also reduced more than $16\%$.
%

%
\textcolor{black}{\begin{table}[htbp]
\begin{center}
\caption{The RMS error by $G$ and $G_{\rm best}$}
 \scalebox{1.3}{ \begin{tabular}{|l|l|l|l|} \hline
     & $G$ &$G_{{\rm best}}$ \\ \hline 
 $B_{f}~~(\rm{MeV})$&$1.39$& --- \\ \hline
 Binding Energy (MeV)&$2.84$& $1.43$ \\ \hline
  $\mathcal{I}_{nn}~~(\rm{MeV}^{-1})$&$1.43$& $1.20$\\ \hline
  \end{tabular}}
     \end{center}
\end{table}}

In the following, we use the linear fit of the $\alpha$ value as shown in Fig.~\ref{Fig5} to calculate the 
systematic properties in the actinide region. 

\begin{figure}[htbp]
\centering\includegraphics[scale=0.5]{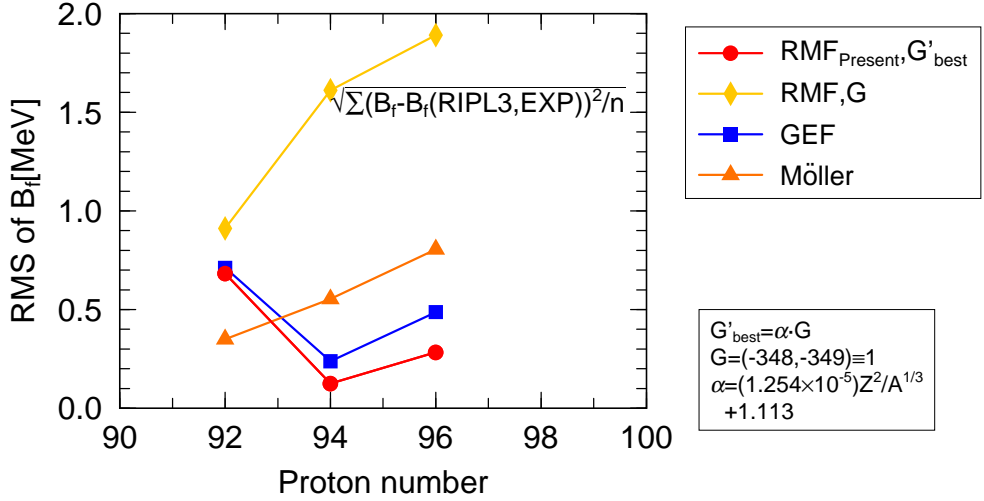}
\caption{RMS for the fission barrier, being RIPL3 taken as 
the standard.}
\label{Fig7}
\end{figure}

\begin{figure}[htbp]
\centering\includegraphics[scale=0.45]{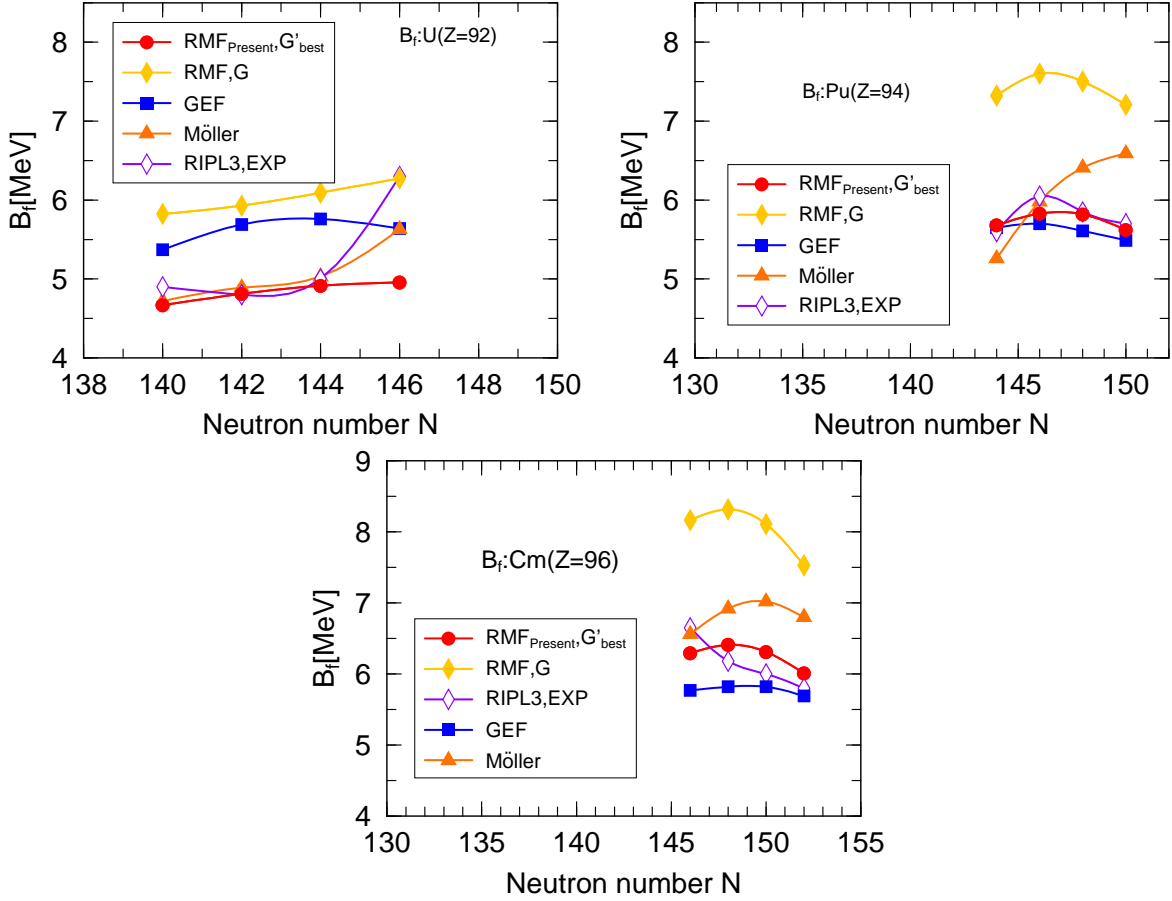}
\caption{Comparison of the calculation results of inner fission barrier by RMF (linear fit to \textcolor{black}{$G'_{\rm best}$}  with RMF($G$), GEF, M$\ddot{{\rm{o}}}$ller and RIPL3.}
\label{Fig8}
\end{figure}

Firstly, \textcolor{black}{we compare RMS values of the inner 
fission barrier height from \textcolor{black}{two} sources (GEF \cite {GEF} and M\"oller \cite {Moller}) and present RMF (original $G$ and \textcolor{black}{$G'_{\rm best} = \alpha \cdot G$ with $\alpha$} given as a linear fit) against the experimental data given in RIPL3 \cite{RIPL3} for $Z=$90, 92 and 94 elements in Fig.~\ref{Fig7}.} What is clear from Fig.~\ref{Fig7} is that M\"oller's RMS of the barrier calculated by the macroscopic-microscopic model and RMF($G$) behaves differently from \textcolor{black}{GEF} and RMF (\textcolor{black}{$G'_{\rm{best}}$}).  The RMF(\textcolor{black}{$G'_{\rm best}$}) result for uranium ($ Z = 92 $) has large RMS due to the inability to reproduce the bouncing behavior of $^{238}{\rm U}$ (see Fig.~\ref{Fig8}). It was found that the presently-determined \textcolor{black}{$G'_{\rm best}$} gives small RMS in an overall manner, especially for Pu and Cm isotopes.

Nextly, the calculated inner fission barrier heights by RMF (\textcolor{black}{$G'_{\rm best}=\alpha \cdot G$ 
where $\alpha$ is given by a linear fit}) and RMF ($G$) are compared for each element in \textcolor{black}{Fig.~\ref{Fig8}}.  It can be seen that for nuclides other than $^{238}{\rm U}$, the agreement with the data (\textcolor{black}{open} purple diamonds) is improved by about 2 MeV by calculations with \textcolor{black}{$G'_{\rm best}$} (red filled circles) than those with $G$ (filled yellow diamonds).  A comparison of the inner fission barrier obtained by using \textcolor{black}{$G'_{\rm best}$} with other literature values ( GEF \cite{GEF} (blue filled squares) and M$\ddot{\rm o}$ller \cite{Moller} (orange filled triangles)) is also shown in this figure.  
We recognize that the present RMF calculation with \textcolor{black}{$G'_{\rm{best}}$} reproduces the fission barrier height quite well compared to other calculations.  
The $B_f$ of $^{238}$U shows a sudden jump from that of $^{236}$U, but this behaviour could not be reproduced by the present parametrization given by a linear function
of $Z^2/A^{1/3}$.  Otherwise, 
the present parametrization shows an overall agreement with experimental data for other isotopes of U and all of
Pu and Cm isotopes.

Changes in the description of the binding energies and moment-of-inertia of pairing rotation $\mathcal{I}_{nn}$ calculated with $G$ and \textcolor{black}{$\alpha \cdot G$} where \textcolor{black}{$\alpha$} 
is given by the linear fit in Fig.~7 is indicated in Tables 3 and 4.  As we see, the RMS errors for the both quantities were reduced noticeably.  It is especially interesting
that improvements in the binding energies are significant, since the RMF parameters were determined originally to reproduce binding energies of lighter nuclei, so
we may expect that there is \textcolor{black}{a} little room to improve description of the binding energies.   
\begin{table*}[htbp]
  \begin{center}
    \begin{tabular}{c}
%
      \hspace*{-1.5cm}
  \begin{minipage}{0.45\hsize}
        \begin{center}
          \caption{RMS(B.E.)\textcolor{black}{[MeV]}}
          \scalebox{1.0}{\begin{tabular}{|l|l|l|l|l|l|} \hline
 Element& $G$&$\textcolor{black}{G'_{{\rm best}}}$ \\ \hline 
 Uranium& $3.450$&$1.934$ \\ \hline
 Plutonium&$3.757$&$2.226$ \\ \hline
 Curium&$3.377$&$1.905$ \\ \hline
          \end{tabular}}
        \end{center}
      \end{minipage} 
      \hspace*{1cm}
      \begin{minipage}{0.45\hsize}
        \begin{center}
          \caption{RMS(\textcolor{black}{$\mathcal{I}_{nn}$})\textcolor{black}{[${\rm{MeV}}^{-1}$]}}
         \scalebox{1.0}{   \begin{tabular}{|l|l|l|l|l|l|} \hline
 Element&  $G$&$\textcolor{black}{G'_{{\rm best}}}$ \\ \hline 
 Uranium& $1.022$&$0.840$ \\ \hline
 Plutonium&$1.182$&$0.756$\\ \hline
 Curium&$2.471$&$1.788$
  \\ \hline
           \end{tabular}}
        \end{center}
      \end{minipage}
    \end{tabular}
  \end{center}
\end{table*}
%



\section{Summary}
We \textcolor{black}{systematically investigated} height of the inner fission barrier of actinide nuclei using BCS pair correlation as a residual interaction in relativistic mean-field theory.  In all of the NLV-20, DD-ME2 and DD-PC1 parameterizations, the inner fission barrier was overestimated by 1 to 2 MeV \textcolor{black}{for $^{240}$Pu}.    
On this basis, the experimental values of the inner fission barrier could be reproduced better by appropriately enhancing the pair correlation force by about 13 \%, and  new systematics of the 
pairing strength was constructed for NLV-20 as a linear fit which can be applicable to nuclei in the 
actinide region. In addition, we brought the concept of the pair rotation, which has been pointed out recently as a method for purely evaluating the pair correlation effects compared to the conventional method, and validated the pairing strength adjusted to reproduce the inner fission barrier. 
As a result, consistent results were obtained in which not only the inner fission barrier heights but also the accuracy of binding energy and the pair rotation moment-of-inertia were improved 
\textcolor{black}{simultaneously}, and this new systematics can be \textcolor{black}{used} for predictions
of various quantities in the actinide region.  Although our results were obtained 
in terms of the 
constant pairing strength in the relativistic mean field theory,  
the same trend can be concluded for a density dependent pairing interaction, or even
for non-relativistic approaches such as Skyrme Hartree-Fock method.  
\textcolor{black}{It must be also pointed out that the inclusion of the \textcolor{black}{triaxiality} will change the 
value of the $\alpha$ parameter given as a linear fit, but the essential conclusions should remain unchanged.}  

\section{Acknowledgement}
The authors are grateful to Prof. J.~Maruhn (Goethe University Frankfurt) for an essential support to carry out this study.


\begin{thebibliography}{100}
\bibitem{review}M.~Bender et al., J.~Phys.~G: Nucl.~Part.~Phys.~{\bf 47}, 113002 (2020).
\bibitem{Bohr}N.~Bohr and J.~A.~Wheeler, Phys.~Rev.~56, 426 (1939).
\bibitem{exp} S.~Bj{\o}rnholm, J.~E.~Lynn, Rev.~Mod.~Phys. 52 (1980) 725.
\bibitem{Baran}A.~Baran and K.~Pomorski, Nucl.~Phy.~A 361 (1981) 83-101
\bibitem{RMFresult}K.~Rutz, J.~A.~Maruhn, P.~G.~Reinhard and W.~Greiner, Nucl. Phys. A 590, 680 (1995).
\bibitem{RMFBCS}S.~Karatzikos, A.~V.~Afanasjev, G.~A.~Lalazissis, P.~Ring, Phys.Lett. B 689 (2010) 72-81
\bibitem{Abu}H.~Abusara, A.~V.~Afansjev, and P.~Ring, Phys.Rev.C 82(2010) 044303(2010)
\bibitem{Agb}S.~E.~Agbemava, A.~V.~Afanasjev, D.~Ray, and P.~Ring, Phys.Rev.C 95(2017) 054324
\bibitem{Shi}Z.~Shi, A.~V.~Afanasjev, Z.~P.~Li, and J.~Meng, Phys.Rev.C 99(2019) 064316
\bibitem{China1}B.~N.~Lu, E.~G.~Zhao, and S.~G.~Zhou, Phys.~Rev.~C 85(2012) 011301(R)
\bibitem{China2}B.~N.~Lu, J.~Zhao, E.~G.~Zhao, and S.~G.~Zhou, Phys.~Rev.~C 89(2014) 014323
\bibitem{China3}J.~Chao, B.~N.~Lu, T.~Niksic, D.~Vretenar, and S.~G.~Zhou, Phys.~Rev.~C 93(2016) 044315
\bibitem{China4}V.~Prassa, T.~Niksic, G.~A.~Lalazissis, and D.~Vretenar, Phys.~Rev.~C 86(2012) 024317
\bibitem{SkyAx}P.-G.~Reinhard, B.~Schuetrumpf, and J.~A.~Maruhn, Comput.~Phys.~Commun 258 (2021) 107603.
\bibitem{tcsm}J.~A.~Maruhn and W.~Greiner, Z.~Phys.~251, 431(1972).
\bibitem{History}A.~Baran, M.~Kowal, P.-G.~Reinhard, L.~M.~Robledo, A.~Staszczak, M.~Warda, Nucl.~Phys.~A 994 (2015) 442-470
\bibitem{Sad}J.~Sadhukham, J.~Dobaczewski, W.~Nazarewicz, J.~A.~Sheikh, and A.~Baran, Phy.~Rev.~C 90 (2014) 061304
\bibitem{Skyrme} K.~Kean, T.~Nishikawa, and Y.~Iwata, JPS Conf.~Proc.~32, 010018 (2020)
\bibitem{Gogny}R.~Rodriguez-Guzman, Y.~M.~Humadi, L.~M.~Robledo, Eur.~Phys.~J. A (2020) 56:43
%
\bibitem{energy}B.~D.~Serot and J.~D.~Walecka, Adv.~Nucl.~Phys.~16 (1986) 1.
\bibitem{pairingrotation}N.~Hinohara and W.~Nazarewicz, Phys.~Rev.~Lett.~116, 152502 (2016).
\bibitem{pairingrotation2}D. M. Brink and R. A. Broglia, Nuclear Superfluidity, Pairing in Finite Systems (Cambridge University Press, Cambridge, England, 2005).
%
\bibitem{RIPL3}R.~Capote, M.~Herman, P.~Oblozinsky, P.~G.~Young, S.~Goriely, T.~Belgya, A.~V.~Ignatyuk, A.J.~Koning, S.~Hilaire, V. A.~Plujko, M.~Avrigeanu, O.~Bersillon, M.~Chadwick, T.~Fukahori, Zhigang Ge, Yinlu Han, S.~Kailas, J.~Kopecky, V.M.~Maslov, G.~Reffo, M.~Sin, E.Sh.~Soukhovitskii, P.~Talou, Nucl.~Data Sheets 110 (2009) 3107.
\bibitem{ALICE}M.~Blann, ALICE-91, Statistical Model Code System with Fission Competition, RSIC CODE PACKAGE PSR-146.
\bibitem{EMPIRE}M.~Herman, R.~Capote, B.~V.~Carlson, P.~Oblozinsky, M.~Sin, A.~Trkov, H.~Wienke, and V.~Zerkin Nucl.~Data Sheets 108 (2007) 2655.
\bibitem{GNASH}P.~G.~Youngand E.~D.~Arthur GNASH; A Preequilibrium, Statistical Nuclear Model Code for Calculation of Cross Sections and Emission Spectra, LA-6947 (1977).
\bibitem{UNF}J.~ZHANG, Nucl.~Sci.~Eng.~142,207(2002).
\bibitem{TALYS}A.~J.~Koning, S.~Hilaire, and M.~C.~Duijvestijn, TALYS- 0.64. A Nuclear Reaction Program. User Manual. NRG
Report 21297/04.62741/P FAIJAK/AK (Dec 5, 2004).
\bibitem{ccone}O.~Iwamoto, N.~Iwamoto, S.~Kunieda, F.~Minato and K.~Shibata, Nuclear Data Sheets 131, 259-288 (2016).
\bibitem{Lagrangian1}S.~Gmuca, Nucl.~Phys.~A {\rm{547}}, 447-458 (1992).
\bibitem{Lagrangian2}Y.~Sugahara and H.~Toki, Nucl.~Phys.~A{\rm{579}}, 557-572 (1994).
\bibitem{Para} K.~Rutz "Struktur von Atomkernen im Relativistic-Mean-Field-Modell", Doctoral Dissertation, Johann Wolfgang Goethe-Universit$\ddot{a}$t in Frankfurt am Main (1999)
\bibitem{program}T.~Niksic,~D.~Vretenar, N.~Paar,~and P.~Ring, Comput.~Phys.~Commun~185~(2014)~1808$-$1821 
\bibitem{bender}M.~Bender, K.~Rutz, P.-G.~Reinhard and J.~A.~Maruhn, Eur.~Phys.~J.~A {\bf 8}, 59-75~(2000).
\bibitem{bindexp}W.J.~Huang, G.~Audi, Meng Wang, F.G.~Kondev, S.~Naimi, Xing Xu, Chinese Physics C Vol. 41, No. 3 (2017) 030002.
%
\bibitem{Nillson}S.G.~Nilsson and I.~Ragnarsson, Shapes and Shells in Nuclear Structure, Cambridge University Press, Cambridge, 1995.
\bibitem{DDME2}G.A.~Lalazissis,~T.~Niksic,~D.~Vretenar, and P.~Ring, Phys.~Rev.~C~${\bm{71}}$,~024312~(2005)
\bibitem{DDPC1}T.~Niksic,~D.~Vretenar and P.~Ring, Phys.~Rev.~C~${\bm{78}}$,~034318~(2008)
\bibitem{GEF}K.-H.~Schmit, B.~Jurado, C.~Amouroux and C.~Schmitt, Nuclear Data Sheets Vol.131 (2016) 107-221.
\bibitem{Moller}P.~Moller, A.~J.~Sierk, T.~Ichikawa, A.~Iwamoto, R.~Bengtsson, H.~Uhrenholt and S.~Aberg, Phys.~Rev.~C 79 (2009) 064304.
\end{thebibliography}
\end{document}